\documentclass[aip,jap,reprint,amsmath,amssymb,longbibliography,footinbib]{revtex4-1}
\usepackage[colorlinks]{hyperref}

\usepackage{graphicx,subfigure,epsfig}
\usepackage{dcolumn}
\usepackage{bm}
\usepackage{undertilde}

\hypersetup{
    colorlinks=true,       
    linkcolor=black,       
    citecolor=black,       
    filecolor=black,       
    urlcolor=black,        
    pdfpagelayout=SinglePage
}

\begin{document}

\title{On the moving contact line singularity}

\author{Rouslan Krechetnikov}

\affiliation{Department of Mathematics, University of Alberta, Edmonton, AB, T6G 2G1, Canada}

\date{April 14, 2018}

\begin{abstract}
Given that contact line between liquid and solid phases can move regardless how negligibly small are the surface roughness, Navier slip, liquid volatility, impurities, deviations from the Newtonian behavior, and other system-dependent parameters, the problem is treated here from the pure hydrodynamical point of view only. In this note, based on straightforward logical considerations, we would like to offer a new idea of how the moving contact line singularity can be resolved and provide support with estimates of the involved physical parameters as well as with an analytical local solution.
\end{abstract}

\maketitle

\section{Basic reasoning}

\begin{figure}[h!]
\includegraphics[width=1.75in]{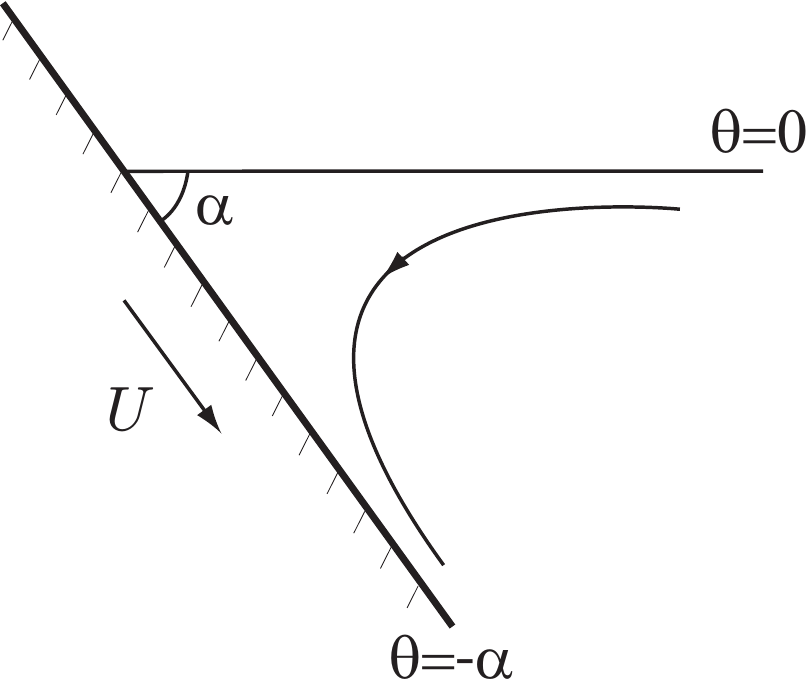}
\caption{Wedge configuration.}\label{fig:wedge}
\end{figure}

Starting with the works of Moffatt \cite{Moffatt:1964} and Huh \& Scriven \cite{Huh:1971}, it is well understood that the flow in the sufficiently close proximity of the moving contact line is in the Stokes regime. In the wedge-geometry configuration, cf. figure \ref{fig:wedge}, the symmetry considerations dictate the unique stream-function solution to the corresponding boundary value problem of the form $\psi = U r f(\theta)$, which exhibits the $r^{-1}$ singularity in the shear stress and pressure thus making both the total force exerted on the solid surface and the rate of viscous dissipation logarithmically divergent. As a result, the fluid particles experience infinite (advective) acceleration when traveling from the interface to the substrate. Various remedies to the difficulty were offered, in particular based either on actual and effective (for rough substrates) slip or existence of a precursor film, which apply to certain liquid-solid combinations. While it is clear that substrate roughness can be reduced down to sub-nanometer scale and there are many situations when neither a precursor film nor a measurable slip at reasonable shear rates are present, the idea that diverging stresses may induce slip or even non-Newtonian behavior in otherwise well-behaving liquids under normal conditions at the very least deserves some estimates, which will be done here for water only.

The distance at which slip may occur is dictated by the balance of intermolecular interactions $O(A/l)$ and viscous forces $O(\mu \, l^{2} \, \dot{\gamma})$, where $\mu$ is the dynamic viscosity, $A \simeq 10^{-19} \, \mathrm{J}$ the Hamaker constant, and $l \simeq 0.3 \, \mathrm{nm}$ a typical molecular length. As a result, the shear rate $\dot{\gamma} \simeq 10^{12} \, \mathrm{s}^{-1}$, from where, for the contact line velocity $U=1 \, \mathrm{mm/s}$, we find the critical distance from the wedge apex $r^{*} \simeq 10^{-15} \, \mathrm{m}$, which is well below $l$ and, in fact, on the order of the hydrogen nucleus diameter. Hence, for a typical situation, one should not rely upon slip to resolve the singularity. Similarly, one might argue that the liquid behaves as non-Newtonian at elevated values of shear stresses. This takes place when the flow time-scale becomes comparable to the relaxation one $\mathcal{T} \sim \mu/K$ below which liquids behave like solids (and hence non-Newtonian effects appear); here $K$ is the bulk modulus typically on the same order as the shear modulus $G \simeq \rho \, c^{2}$, with $c$ being the speed of sound and $\rho$ the density. Hence, for water we find $\mathcal{T} \simeq 10^{-12} \, \mathrm{s}$. Comparing with $\dot{\gamma}^{-1} \simeq r^{*} / U$ for the same value of $U$ we again get $r^{*} \simeq 10^{-15} \, \mathrm{m}$.

\begin{figure}[h!]
\includegraphics[width=2.0in]{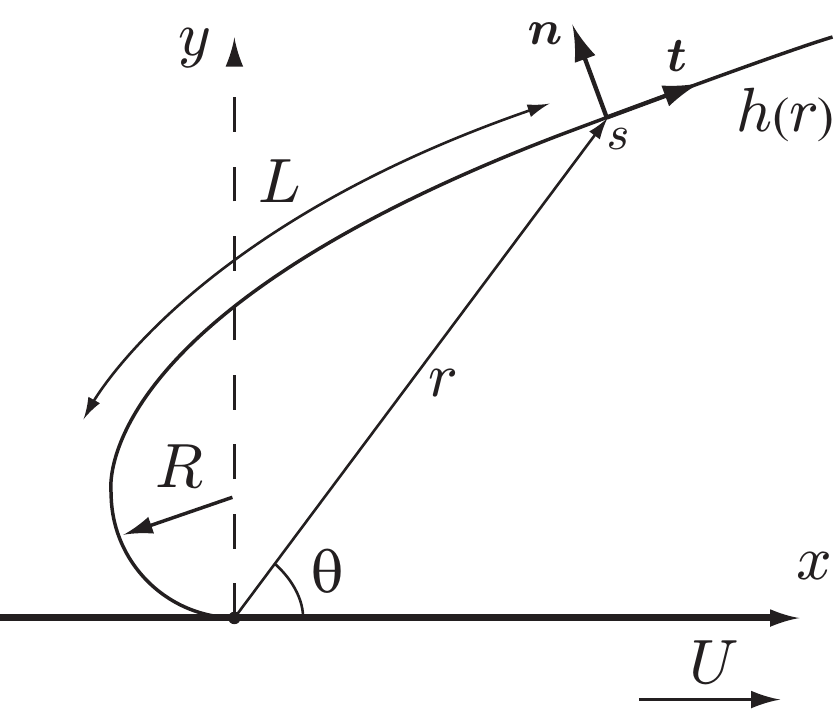}
\caption{On scalings.}\label{fig:scalings}
\end{figure}
Thus, we are left with water sticking to the substrate while its contact line is still capable of moving. The only conceivable logical conclusion within the realms of classical hydrodynamics is that the singularity is an artefact of the assumed sharp wedge geometry. Because Nature does not allow unresolved singularities, in such a situation the next reasonable step is to abandon this geometrical setting, i.e. to assume that the wedge apex is not perfectly sharp and the liquid must meet the interface at the angle $\theta=\pi$, in which case the stresses and pressure previously calculated \cite{Moffatt:1964,Huh:1971} vanish and thus  the contact line singularity is removed. Given that the apparent macroscopic contact angle $\alpha$ is generally different from $\pi$, the only way to reconcile it with the microscopic contact angle $\pi$ is via introducing a highly curved region as in figure \ref{fig:scalings}, which must be very small given that it is not commonly observed at the macroscopic level. The natural candidate for rounding off the tip is surface tension $\sigma$ introducing some small radius of curvature $R$, which can be found by balancing viscous stresses and capillary pressure producing $R \simeq r \, \sigma / (\mu \, U)$, i.e. within this model $R$ is not uniquely determined and hence there could be a host of solutions of the type shown in figure \ref{fig:scalings}. Hence, one needs some extra physical effect to settle $R$ uniquely, which is the subject of the subsequent discussion.

With the above introduced logical constrains, the only inevitable effect which we have at our disposal and which is inherently present, but, to the author's knowledge, has not been taken into account, is the heat produced by viscous dissipation in the region of high curvature $R^{-1}$ and thus of significant stresses. But the heat will naturally lead to surface tension variations and hence Marangoni stresses, thereby providing a potential mechanism for establishing the length scale $R$.

\section{Scaling} \label{sec:scaling}

To support the idea that the contact line singularity can be resolved with surface tension variations, let us perform simple scaling analysis for the situation considered by Moffatt \cite{Moffatt:1964} and Huh \& Scriven \cite{Huh:1971}, i.e. when the contact line motion is established and thus steady-state. Adopting the polar coordinate system, cf. figure \eqref{fig:wedge}, the general dynamic conditions at the interface $\theta=h(r)$ read:
\begin{subequations}
\label{dynamic}
\begin{align}
\label{dynamic:normal}
p - \frac{2 \, \mu}{r^{3}} \, \frac{\left(v_{\theta}+u\right) - r \, h_{r} \, \left(u_{\theta} + r v_{r} - v\right)}{r^{-2}+h_{r}^{2}} = \sigma \, \nabla \cdot \mathbf{n}, \\
\label{dynamic:tangent}
\scalebox{1.25}[1.25]{$\frac{\left(\frac{1}{r^{2}} - h_{r}^{2}\right) \left(\frac{u_{\theta}}{r}+v_{r}-\frac{v}{r}\right) + \frac{2 h_{r}}{r} \left(\frac{v_{\theta}}{r}+\frac{u}{r}-u_{r}\right)}{r^{-2}+h_{r}^{2}} = \frac{\sigma_{s}}{\mu}$},
\end{align}
\end{subequations}
where $u$ and $v$ and the $r$- and $\theta$-components of the velocity field, $\nabla \cdot \mathbf{n}$ the interfacial curvature, which is positive for the configuration shown in figure \ref{fig:scalings}, and $s$ the arclength. From the tangential stress balance
\begin{align}
\frac{\mu}{r} \frac{\partial u}{\partial \theta} \simeq \frac{\mathrm{d} \sigma}{\mathrm{d} T} \frac{\partial T}{\partial s}
\end{align}
at the free interface under the assumption of linear dependence of surface tension on temperature $\sigma = \sigma_{0} - \gamma \, \Delta T$ (with $\gamma >0$ for normal substances) we find
\begin{align}
\label{balance:force}
\mu \frac{U}{L} \sim \gamma \frac{\Delta T}{L} \ \Rightarrow \ \Delta T \sim \frac{\mu \, U}{\gamma},
\end{align}
which implies that the viscous stress in the ``square'' $L \times L$ region balances the Marangoni one. The energy balance equation for an incompressible flow is
\begin{align}
\label{eqn:energy}
\rho \, c_{p} \, \mathbf{v} \cdot \nabla T = \kappa \, \nabla^{2} T + \frac{\mu}{2} \, \epsilon_{ik} \, \epsilon^{ik},
\end{align}
where $\epsilon_{ik}$ is the strain rate tensor. Since heat conduction alone does not lead to a finite temperature distribution in the presence of a point heat source (the last term in the energy equation), the only way to get a finite length scale $L$ for the temperature variation $\Delta T$ is by balancing the advection and conduction terms
\begin{align}
\label{balance:energy-distribution}
\rho \, c_{p} \, U \frac{\Delta T}{L} \sim \kappa \, \frac{\Delta T}{L^{2}} \ \Rightarrow \ L = \frac{\kappa}{U \, \rho \, c_{p}}.
\end{align}
It is at this spatial scale when the advection effects equilibrate with the diffusion in \eqref{eqn:energy}. On shorter length scales the diffusion is dominant and in this two-dimensional problem leads to a logarithmic temperature distribution (i.e. the fundamental solution of the Laplace equation with a point source), which does not set up any characteristic length scale. On longer length scales the temperature distribution is leveled exponentially due to the advection as follows from equation \eqref{eqn:energy}. The radius of curvature $R$ of the singular region is then found from the balance of heat energy production in this region and the subsequent transfer over the larger region of length $L$, cf. figure \ref{fig:scalings}:
\begin{align}
\label{balance:energy-production}
\mu \left(\frac{U}{R}\right)^{2} \sim \kappa \, \frac{\Delta T}{L^{2}},
\end{align}
i.e. indeed the last term in equation \eqref{eqn:energy} acts as a point source provided $R \ll L$. From (\ref{balance:force},\ref{balance:energy-distribution},\ref{balance:energy-production}) we arrive at
\begin{align}
\label{scalings}
L \sim \frac{\kappa}{U \, \rho \, c_{p}}, \ \Delta T \sim \frac{\mu \, U}{\gamma}, \ R \sim \frac{1}{\rho \, c_{p}} \sqrt{\frac{\gamma \, \kappa}{U}}.
\end{align}
Taking the parameter values for water ($\gamma = 1.4 \cdot 10^{-4} \, \mathrm{N/(m \cdot K)}$, $\kappa = 0.56 \, \mathrm{W/(m \cdot K)}$, $c_{p} = 4185.5 \, \mathrm{J/(kg \cdot K)}$, $\rho = 10^{3} \, \mathrm{kg/m^{3}}$, $\mu = 1.002 \, \mathrm{mPa \cdot s}$, $U = 10^{-3} \, \mathrm{m/s}$), we estimate
\begin{align}
\label{estimates}
L \simeq 0.1 \, \mathrm{mm}, \ R \simeq 0.1 \, \mathrm{\mu m}, \ \Delta T \simeq 10^{-2} \, \mathrm{K}.
\end{align}
A few comments are in order. The resulting value of $R$ is well above the Tolman length and other nanometric effects, i.e. only ordinary fluid physics is required to resolve the singularity. The convective acceleration is estimated as $U^{2}/R = O(10) \, \mathrm{m/s^{2}}$, being far from unreasonable. While the temperature variation seems to be small, its gradient is $O(10^{2}) \, \mathrm{K/m}$, which is an order of magnitude higher than the one required for the onset of Rayleigh-Benard convection in a water layer of $1 \, \mathrm{cm}$ thickness. The temperature variation $\Delta T$ is also comparable with the molecular dynamics estimate: namely, even without taking into account other heat transfer effects, when a water molecule with the average velocity of $\langle v \rangle \simeq 500 \, \mathrm{m/s}$ bounces from the wedge apex moving with velocity $U$ it acquires the kinetic energy $\sim 2 \, \langle v \rangle \, U$ per molecule mass, which, when compared to the average kinetic energy at room temperature, yields $\Delta T \simeq 2 \cdot 10^{-3} \, \mathrm{K}$. The latter independent estimate again highlights the purely formal nature of the moving contact line singularity (which predicts infinite $\Delta T$ due to the diverging rate of energy dissipation) resulting from the deficiency of the wedge model in figure \ref{fig:wedge}.

\section{Wedge problem with Marangoni effects}

Paradoxically, despite the deficiency of the wedge model \cite{Moffatt:1964,Huh:1971}, we can gain a few further important insights into the problem at hand by revisiting the Moffatt's solution \cite{Moffatt:1964} with inclusion of Marangoni stresses. With reference to figure \ref{fig:wedge}, it is clear that the solution of the corresponding Stokes problem is still of the form $\psi = U \, r \, f(\theta)$, so that the velocity components are given by
\begin{align}
u = \frac{1}{r} \frac{\partial \psi}{\partial \theta} = U \, f^{\prime}(\theta), \ v = - \frac{\partial \psi}{\partial r} = - U \, f(\theta).
\end{align}
The appropriate boundary conditions read:
\begin{subequations}
\begin{align}
\theta = -\alpha&: \ u = U, \ v = 0, \\
\theta = 0&: \ v = 0, \ \mu \frac{1}{r} \frac{\partial u}{\partial \theta} = \frac{\partial \sigma}{\partial r} > 0,
\end{align}
\end{subequations}
or, in the stream function form:
\begin{subequations}
\begin{align}
\theta = -\alpha&: \ f(-\alpha) = 0, \ f^{\prime}(-\alpha) = 1, \\
\label{BC:tangent}
\theta = 0&: f(0) = 0, \ f^{\prime\prime}(0) = \frac{1}{Ca} \, r \frac{\partial \sigma}{\partial r} \ \equiv \tau,
\end{align}
\end{subequations}
where the expression in the tangential balance is non-dimensionalized for convenience with $r \rightarrow L \, r$ with $L$ from the previous section; here $Ca = \mu \, U / \sigma_{0}$ is the capillary number. Since
\begin{align}
\frac{\partial \sigma}{\partial r} = \frac{\mathrm{d} \sigma}{\mathrm{d} T} \frac{\partial T}{\partial r},
\end{align}
the only way the boundary condition \eqref{BC:tangent} can be satisfied is if $\Delta T \sim \ln{r}$ (for the assumed linear dependence of surface tension on temperature), which effectively decouples the velocity field from the energy equation \eqref{eqn:energy} -- such a natural occurrence of the logarithmic solution is consonant with the conclusion in previous section. Note that since, based on physical considerations, $\partial \sigma / \partial r > 0$ ($\tau>0$) and $\mathrm{d} \sigma / \mathrm{d} T < 0$, then $\partial T / \partial r < 0$ and thus $\Delta T \sim \ln{r}$ must have a negative sign (corresponding to a decrease in temperature), since the argument $r < 1$ after non-dimensionalization by $L$.

\begin{figure}
\centering
\epsfig{figure=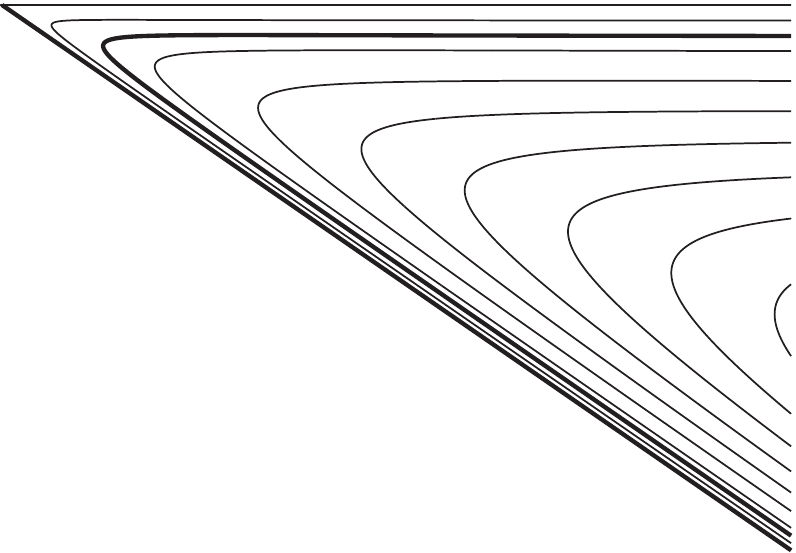,height=1.5in}
\caption{Flow field in the wedge in the same geometry as in figure \ref{fig:wedge}. The bold streamline depicts the one with the maximum radius of curvature $R$ -- an approximate solution embedded in the modified Moffatt's solution.}\label{fig:moffatt}
\end{figure}
The resulting solution for the velocity field takes the form:
\begin{multline}
\label{f1:general}
f(\theta) = \frac{\theta \cos{\theta} \sin{\alpha}-\alpha \cos{\alpha} \sin{\theta}} {\sin{\alpha} \cos{\alpha} - \alpha} + \\ \frac{\tau}{2} \, \frac{\theta \sin{\alpha} \sin{(\alpha+\theta)} - \alpha (\alpha + \theta) \sin{\theta}} {\sin{\alpha} \cos{\alpha} - \alpha},
\end{multline}
where the first term is the same as in Moffatt's solution \cite{Moffatt:1964}, while the latter term is due to the Marangoni contribution. The free surface velocity is then independent of $r$ and reads
\begin{align}
\label{us:general}
u_{s} &= U \, f^{\prime}(0) = \\ &- U \left[1 - \frac{(\alpha - \sin{\alpha})(1 + \cos{\alpha})} {\alpha - \sin{\alpha} \cos{\alpha}} - \frac{\tau}{2} \, \frac{\alpha^{2} - \sin^{2}{\alpha}} {\alpha - \sin{\alpha} \cos{\alpha}}\right], \nonumber
\end{align}
where the second term is positive
\begin{align}
\frac{(\alpha - \sin{\alpha})(1 + \cos{\alpha})} {\alpha - \sin{\alpha} \cos{\alpha}} > 0
\end{align}
with the maximum of $1/2$ at $\alpha=0$. The last term in \eqref{us:general} is non-negative
\begin{align}
\frac{\alpha^{2} - \sin^{2}{\alpha}} {\alpha - \sin{\alpha} \cos{\alpha}} \ge 0
\end{align}
and vanishes at $\alpha=0$. Since $\tau > 0$, the Marangoni stress partially suppresses the otherwise significant jump of velocity from $-U/2$ at the interface to $U$ at the substrate. In fact, if $\tau$ is large enough, $u_{s}$ changes sign at the critical shear:
\begin{align}
\label{tau:star}
\tau^{*} = - 2 \, \frac{\alpha \cos{\alpha} - \sin{\alpha}}{\alpha^{2}-\sin^{2}{\alpha}} > 0,
\end{align}
above which there appears a secondary recirculating flow. Notably, $\tau^{*} \rightarrow \infty$ as $\alpha \rightarrow 0$ (which makes sense as the interface is closer to the substrate), but $\tau^{*} \rightarrow 2/\pi$ as $\alpha \rightarrow \pi$.

For contrast, let us remind a few key details about Moffatt's solution \eqref{f1:general} with $\tau=0$. In the limit when $\alpha=\pi$ so that the interface meets the solid tangentially as shown in figure \ref{fig:scalings}, the solution \eqref{f1:general} reduces to
\begin{align}
\label{f:Moffatt-pi}
f(\theta) = -\sin{\theta},
\end{align}
in which case the interfacial velocity $u_{s} = - U$, i.e. there is no discontinuity in velocity when a fluid particle leaves the interface for solid and thus no infinite accelerations are involved. The viscous stresses corresponding to the solution \eqref{f:Moffatt-pi} all vanish identically in the neighborhood of the contact point, $e_{rr} = e_{r\theta} = e_{\theta\theta} = 0$, and thus the singularity is removed.

In the presence of Marangoni stresses, the solution \eqref{f:Moffatt-pi} at $\alpha=\pi$ generalizes to
\begin{align}
f(\theta) = - \sin{\theta} + \frac{\tau}{2} (\pi+\theta) \, \sin{\theta}.
\end{align}
Notably, as opposed to the Moffatt's case, one of the viscous stresses as well as pressure gradients do not vanish at the point of contact even when the liquid interface meets the substrate at $\alpha=\pi$:
\begin{align}
\label{stress:singular}
e_{r\theta} = \tau \frac{\cos{\theta}}{r}, \ \frac{\partial p}{\partial r} = - \tau \frac{\sin{\theta}}{r^{2}}, \ \frac{\partial p}{\partial \theta} = \tau \frac{\cos{\theta}}{r}.
\end{align}
This means, again using the principle that Nature does not sustain unresolved singularities, that at this point the temperature must have an extremum, i.e. $T_{r}=0$. Given the definition of $\tau$ in \eqref{BC:tangent}, $\tau$ vanishes faster than $r$ with $r \rightarrow 0$ and therefore, no singularity is present at the contact line. Another way to look at this is first to start from the situation of a singular stress \eqref{stress:singular}, which inevitably implies an extremum of viscous dissipation and hence $T_{r}=0$ at $r=0$, thus leading to the contradiction that singular stresses can exist. This self-regulated effect allows for the existence of the solution in figure \ref{fig:scalings}.

Another contributing effect that leads to the solution in figure \ref{fig:scalings} is bending of the interface, the mechanism behind which can also be clarified with  the help of the modified Moffatt's solution \eqref{f1:general}. In the absence of surface tension gradients, the tangential stress at the free interface vanishes as in Moffatt's solution for a wedge. This can also be seen from \eqref{dynamic:tangent} after linearization for small $h_{r}$ leading to $u_{\theta}=0$. If initially there is no interfacial curvature, $\nabla \cdot \mathbf{n}=0$, as in the wedge geometry in figure \ref{fig:wedge}, the presence of surface tension gradients leads to the viscous stress deviating from zero $u_{\theta} \sim \sigma_{s} > 0$ and thus $f^{\prime\prime}(0)>0$. Then, as per linearization of \eqref{dynamic:normal}, the pressure must increase away from the constant atmospheric $p_{a}=0$:
\begin{align}
p \sim - \frac{\mu \, h_{r}}{r^{2}} (u_{\theta}+r \, v_{r}-v) = - \frac{h_{r}}{r^{3}}\left[f+f^{\prime\prime}\right]_{\theta=0} > 0,
\end{align}
since $f^{\prime\prime}(0)>0$ and $f(0)=0$, which leads to bulging of the interface due to being unbalanced by the atmospheric pressure. The bulging continues until the capillary pressure  term $\sigma \, \nabla \cdot \mathbf{n}$ balances the viscous stresses in \eqref{dynamic:normal}.

In summary, both effects -- extremal dissipation at the contact line and bending of the interface -- discussed above provide a mechanism for the formation of the steady-state solution in figure \ref{fig:scalings}. Adopting this point of view of a time-transient process enables a proper transition from the static contact line ($U=0$) to its motion at a macroscopic speed $U$ considered here and also explains the distinguished limit nature of scalings \eqref{scalings} when $U \rightarrow 0$: if one starts with the sharp wedge geometry ($R=0$) in the static case, $R$ dynamically increases to the value estimated in \eqref{estimates} without diverging to infinity if one were to plug $U=0$ in formula \eqref{scalings} for $R$.

The streamline $\psi(r,\theta) = r \, f(\theta)$, the radius of curvature of which assumes the value of $R$ estimated above (cf. bold streamline in figure \ref{fig:moffatt} as an illustration), for many practical purposes can be considered as a reasonable local approximation to the solution at the distances $r \ll L$. Indeed, it approaches the substrate as
\begin{align}
\theta + \alpha \sim R/r \rightarrow 0 \ \text{as} \ r \rightarrow \infty,
\end{align}
and the substrate velocity as
\begin{align}
u = 1 + O\left(R/r\right) \rightarrow 1 \ \text{as} \ r \rightarrow \infty,
\end{align}
with the streamfunction $\psi$ assuming the value of $O(10^{-10}) \, \mathrm{m^{2}/s}$ (versus $\psi=0$ at the singular wedge-shaped interface). Given the small value of $R$, the corresponding streamline, which is constant (as the steady state interface streamline continuing along the substrate with the same value), and the flow field interior to it provide a non-singular approximation to the local solution.

\section{Further comments}

While here we discussed the advancing contact angle problem only, the receding one is almost identical to the dip-coating problem of Landau and Levich \cite{Landau:1942}, which predicts $h_{\infty} \sim l_{c} \, Ca^{2/3}$ for the thickness of the film deposited on a substrate moving with velocity $U$ in the direction opposite to gravity. In our case, instead of the gravity-induced capillary length scale $l_{c} =\sqrt{\sigma/(\rho \, g)}$, which is irrelevant near the contact line, the characteristic length $L$ is dictated by the Marangoni force \eqref{balance:energy-distribution} producing $h_{\infty} \sim L \, Ca^{2/3} = O(10^{-8}-10^{-7}) \, \mathrm{m}$ for the conditions considered in this note.

The reader may deem the resolution offered here as a hypothesis, which can be verified only by direct measurements requiring significant accuracy, in particular to resolve relatively small temperature variations on the short length scales \eqref{estimates} and the true submicroscopic interfacial shape near the contact line. This is conceptually no different than what a numerical simulation approximating this intrinsically multi-scale problem may suggest until confirmed by an experiment. Instead, the effort here was put into gaining the physical understanding based on the structure of the governing equations as well as the physical laws we have at our disposal.

\bibliographystyle{unsrtnat}

\end{document}